\documentstyle[aps,manuscript,epsf,multicol]{revtex}

\begin{document}
\draft
%%%%%%%%%%%%%%%%%%%%%%%%%%%%%%%%%%%%%%%%%%
\title{Signal estimation and threshold optimization using an array of
  bithreshold elements}
\author{Aki-Hiro Sato}
\address{Department of Applied Mathematics and Physics, Graduate
  School of Informatics, \\ Kyoto University, Kyoto 606-8501, Japan}
\author{Michihito Ueda}
\address{Matsushita Electric Industrial Co. Ltd. \\ 3-4, Hikaridai,
  Soraku-gun, Kyoto 619-0237, Japan }
\author{Toyonori Munakata}
\address{Department of Applied Mathematics and Physics, Graduate
  School of Informatics, \\ Kyoto University, Kyoto 606-8501, Japan}
%%%%%%%%%%%%%%%%%%%%%%%%%%%%%%%%%%%%%%%%%%
\maketitle
%%%%%%%%%%%%%%%%%%%%%%%%%%%%%%%%%%%%%%%%%%
\begin{abstract}
We consider the problem of optimizing signal transmission through 
multi-channel noisy devices. We investigate an array of bithreshold
noisy devices which are connected in parallel and convergent on a summing
center. Utilizing the concept of noise-induced linearization we
derive an analytical approximation of the normalized power norm and clarify
the relation between the optimum threshold and the standard deviation of 
noises. We show that the optimum threshold value is 0.63 times the
standard deviation of the noises. This relation is applicable to both
subthreshold and suprathreshold inputs.
\end{abstract}
%%%%%%%%%%%%%%%%%%%%%%%%%%%%%%%%%%%%%%%%%%
%%%%%%%%%%%%%%%%%%%%%%%%%%%%%%%%%%%%%%%%%%
%\begin{multicols}{2}
\section{Introduction}
Stochastic resonance (SR) has attracted considerable attention of 
many researchers during the last quarter
century~\cite{Benzi:81,McNamara:89,Wiesenfeld:94,Eichwald:97,Gammaitoni:98,Stocks,Duan:03}.
At first SR was proposed to explain the observed periodicities in 
global climate dynamics~\cite{Benzi:81}. SR occurs when the
signal-to-noise ratio (SNR) for the response of a single nonlinear system to a 
subthreshold sinusoidal input signal has its maximum at a nonzero
noise strength $D$. As well-known, the SR effect is understood as
enhancement of the system input to a subthreshold input signal by
addition of noise.

There are many studies on SR for a single element. For
example Gammaitoni {\it et al.} showed with SNR that a subthreshold
sinusoidal signal to a single threshold element is optimally
transduced by appropriate additive noise~\cite{Gammaitoni}. Collins
{\it et al.} also reported that a single neuron can optimally 
transmit a slowly varying subthreshold aperiodic signal with the aid of
appropriate additive noise~\cite{Collins:95-a}. They proposed the
power norm $C_0$ and the normalized power norm $C_1$ in   
order to measure a correlation between the input signal and the output
signal and showed that both $C_0$ and $C_1$ monmonotonically vary with
increasing the noise strength. It is known as {\it aperiodic
stochastic resonance} (ASR).

Many researchers have both experimentally and numerically studied the
symmetrical stochastic resonator such as the Schmitt
trigger~\cite{Marchesoni:98,Rowe:99,Apostolico:97,Litong:01}. In
recent years, the central attention of SR seems to move to a
network of the stochastic resonators, instead of a single stochastic
resonator, such as the global coupled networks and linear
chains~\cite{Marchesoni:96,Lindner:01}. In more recent years a
parallel array of nonlinear elements gathers a lot of attention, where the
parallel array means that the nonlinear elements are connected in
parallel and convergent on a summing
center~\cite{Stocks,Collins:95-b,Chialvo:97}.

Recently Stocks studied the parallel array of the nonlinear 
devices and reported that the suprathreshold stochastic resonance (SSR) can
be observed on these array motivated by applications to 
signal processing~\cite{Stocks}. Also from the standpoint of
neurophysiology Collins {\it et al.} and Chialvo {\it et al.} studied
an parallel array of noisy neurons can exhibit ASR for 
slowly varying signal~\cite{Collins:95-b,Chialvo:97}. Consequently it
is important to consider the parallel array in both signal processing
and neurophysiology.

The main focus of the article is to find an optimal threshold to transmit
an arbitrary signal on the parallel array of the bithreshold elements.
According to the assumption that the amplitude of the input
signal is smaller than the standard deviation of the noises we derive
an analytical approximation of the normalized power norm. Under this
assumption it is not necessary to distinguish between
subthreshold and suprathreshold. Furthermore we apply the
linear response theory to the system that we consider in the
article. The fundamental idea is the noise-induced linearization,
which is an effect that an ensemble average of output from a nonlinear
system is linearized due to noise~\cite{Dykman:94}. From the
theoretical approximation of the normalized power norm we show that
there exists the optimal threshold to maximize it. 

In fact Stocks and Mannella numerically showed that for a summing
network of FitzHugh-Nagumo equations adjusting the threshold to
maximize information transmission does not remove SR effects. They
pointed out that there is a optimal threshold to maximize the mutual
information~\cite{Stocks:01}. Our result is consistent with their
indication.

The article is organized as follows. In Sec. \ref{sec:model}
we show the array of bithreshold units. It is well-known that Schmitt
trigger is a prototype of bithreshold devices~\cite{Marchesoni:98}. In
Sec. \ref{sec:theory} we theoretically derive an approximation of the
normalized power norm under the assumption that the norm of the input
signal is smaller than the additive noise. Utilizing the approximation
we find an optimal threshold where the normalized power norm is maximized. In
Sec. \ref{sec:simulation} we perform numerical simulations for the
model and show that the threshold value at the maximum normalized
power norm depends on the variance of the additive
noises. Sec. \ref{sec:conclusion} is devoted to concluding remarks.
 
\section{Model}
\label{sec:model}
\begin{figure}[htb]
\centering
\epsfxsize=250pt
\epsfbox{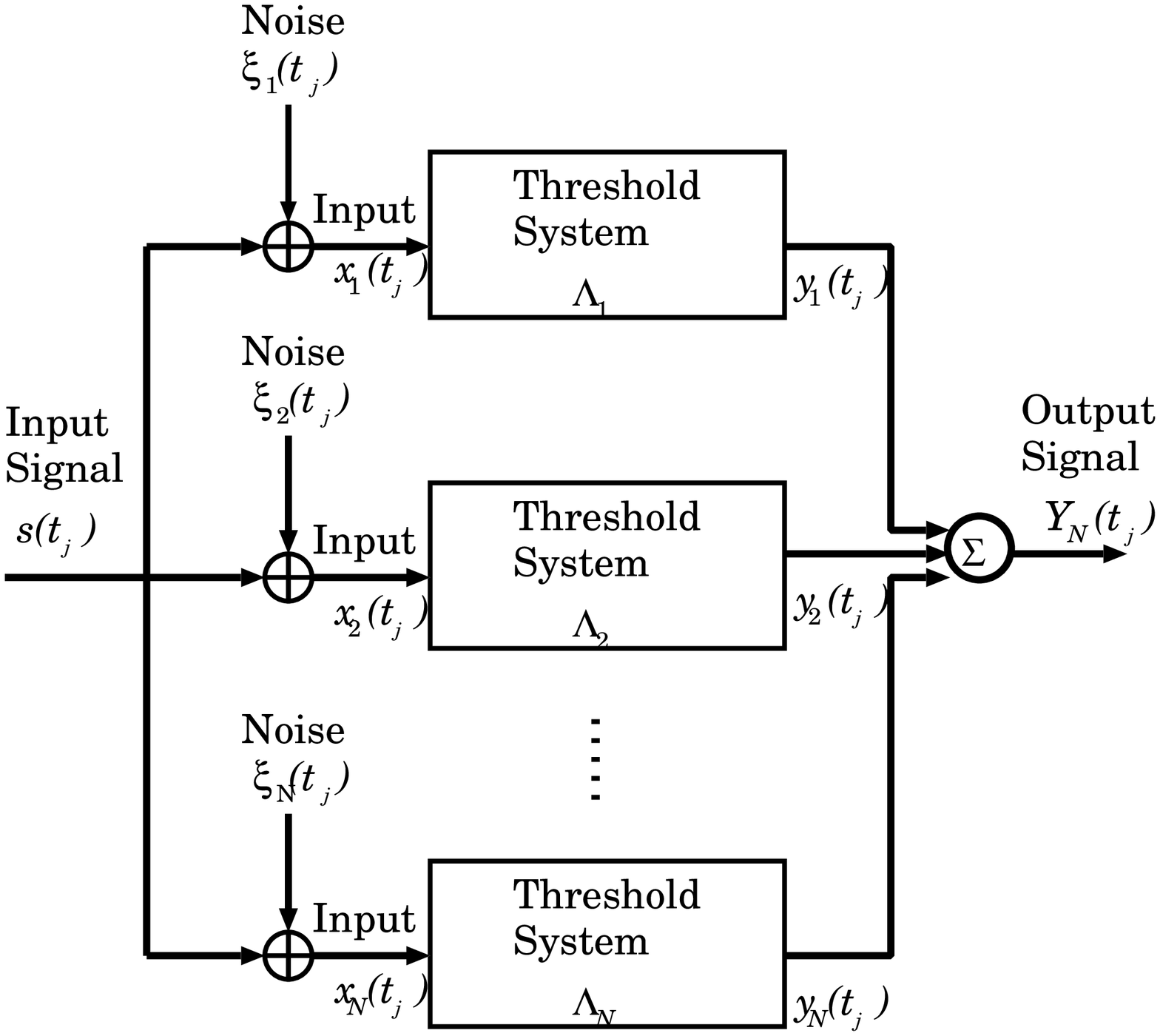}
\caption{The array of bithreshold elements with a summing
  center. $s(t_j)$ exhibits the input of the system. Each subsystem is
  a bithreshold element, which has three output values. $x_i(t_j)$,
  $y_i(t_j)$ and $\Lambda_i$ represent the input, the output and the
  threshold value in the $i$th subsystem, respectively. All the output 
  of subsystems are summed by the summing center and divided by
  $N$. $Y_N(t_j)$ shows the system output.}  
\label{fig:N-array}
\end{figure}

Figure \ref{fig:N-array} displays noisy bithreshold elements connected in
parallel and convergent on a summing center. $s(t_j)$, which is
sampled with a sampling period $T$, namely, $t_j = j T$ ($j = 0, 1,
2, \ldots$), represents a weak aperiodic signal fluctuating around
0 with $x_i(t_j)$ and $y_i(t_j)$ denoting the input to and output from
the $i$th subsystem, respectively. The input to the bithreshold element
is transmitted over noisy channel. Hence the input to the subsystem
is expressed as
\begin{equation}
x_i(t_j) = s(t_j) + \xi_i(t_j),
\end{equation}
where $\xi_i(t_j)$ ($j = 0, 1, 2, \ldots$) is independently sampled
from the Gaussian distribution,
\begin{equation}
p_i(\xi) = \frac{1}{\sqrt{2\pi D_i}} \exp\Bigl(-\frac{\xi^2}{2D_i}\Bigr),
\label{eq:gauss}
\end{equation}
where $D_i>0$ are the variance of $\xi_i(t_j)$. Each bithreshold
element is symmetric and has three output values. It is formalized by  
\begin{equation} 
y_i(t_j) = \left\{
\begin{array}{ll}
1 & \mbox{($x_i(t_j) > \Lambda_i$)} \\
0 & \mbox{($-\Lambda_i \leq x_i(t_j) \leq \Lambda_i$)} \\
-1 & \mbox{($x_i(t_j) < -\Lambda_i$)} 
\end{array}
\right.,
\end{equation} 
where $\Lambda_i>0$ are threshold values. 

The system output through the summing center $Y_N(t_j)$ is defined as, 
\begin{equation}
Y_N(t_j) = \frac{1}{N}\sum_{i=1}^N y_i(t_j),
\label{eq:system-output}
\end{equation}
where $N$ is the number of the subsystems. Without noises, each
input $x_i(t_j)$ cannot cross the threshold value, leading to the system
output $Y_N(t_j) = 0$. With noises having an appropriate
variance, $x_i(t_j)$ can cross the threshold value. 

\section{Theoretical analysis}
\label{sec:theory}
We consider the $i$th bithreshold subsystem. Let $P_+(t_j)$, $P_-(t_j)$
and $P_0(t_j)$ be probabilities that $y_i(t_j)$ takes $1$,
$-1$ and $0$, respectively. For an arbitrary input signal $s(t_j)$
these probabilities are given by,  
\begin{eqnarray}
P_+(t_j) &=&
\frac{1}{2}\mbox{erfc}\Bigl(\frac{\Lambda_i-s(t_j)}{\sqrt{2D_i}}\Bigr), 
\label{eq:probability1}
\\
P_-(t_j) &=&
\frac{1}{2}\mbox{erfc}\Bigl(\frac{\Lambda_i+s(t_j)}{\sqrt{2D_i}}\Bigr), 
\label{eq:probability2}
\\
P_0(t_j) &=& 1 - P_+(t_j) - P_-(t_j),
\label{eq:probability3}
\end{eqnarray}
where $\mbox{erfc}(x)$ is the complementary error function, which is
defined as,
\begin{equation}
\mbox{erfc}(x) = \frac{2}{\sqrt{\pi}}\int_{x}^{\infty}e^{-u^2}du.
\end{equation}

For simplicity we set $\Lambda_i = \Lambda$ and $D_i = D$ for all the
subsystems. From Eqs. (\ref{eq:probability1}), (\ref{eq:probability2})
and (\ref{eq:probability3}) we introduce $\lambda =
\frac{\Lambda}{\sqrt{D}}$ and $\zeta(t_j) =
\frac{s(t_j)}{\sqrt{D}}$. The input signal is subthreshold when
$\zeta(t_j) < \lambda$ and suprathreshold when $\zeta(t_j) > \lambda$.

From Eqs. (\ref{eq:system-output}), (\ref{eq:probability1}),
(\ref{eq:probability2}) and (\ref{eq:probability3}) the ensemble
average of $Y_N(t_j)$ is calculated as
\begin{equation}
\langle Y_N(t_j) \rangle =
\frac{1}{2}\Bigl\{\mbox{erfc}\bigl(\frac{\lambda-\zeta(t_j)}{\sqrt{2}}\bigr) -\mbox{erfc}\bigl(\frac{\lambda+\zeta(t_j)}{\sqrt{2}}\bigr)\Bigr\}.
\label{eq:ensemble}
\end{equation}
Taylor expansion of Eq. (\ref{eq:ensemble}) around $\zeta(t_j)=0$ yields
\begin{equation}
\langle Y_N(t_j) \rangle \approx G(\lambda) \zeta(t_j) 
+ O\Bigl(\zeta(t_j)^3\Bigr),
\label{eq:linearization}
\end{equation}
where $G(\lambda)$ represents the first order coefficient, which is
given by
\begin{equation}
G(\lambda) = \frac{\partial}{\partial \zeta}\langle Y_N(t_j)
 \rangle|_{\zeta(t_j)=0} = \sqrt{\frac{2}{\pi}}
e^{-\frac{\lambda^2}{2}}.
\label{eq:G}
\end{equation}
Thus Eq. (\ref{eq:linearization}) shows that $\langle Y_N(t_j) \rangle$
is a linear function of the input signal $\zeta(t_j)$ for $\langle
|\zeta| \rangle \ll 1$, where $\langle |\zeta| \rangle$ is the norm of
the input signal, measured by the average of the amplitude of the signal.
We call $G(\lambda)$ in Eq. (\ref{eq:linearization}) ``gain''.

Now in order to measure correlation between the input signal $s(t_j)$
and the output signal $Y_N(t_j)$ we introduce the normalized power
norm~\cite{Collins:95-a} 
\begin{equation}
C_1 = \frac{C_0}{[{\overline{s(t_j)^2}}]^{1/2}[\overline{(Y_N(t_j) -
      \overline{Y_N(t_j)})^2}]^{1/2}},
\label{eq:def-C1}
\end{equation}
where $C_0$ is defined as
\begin{equation}
C_0 = \overline{s(t_j) Y_N(t_j)},
\label{eq:power-norm}
\end{equation}
with the overbar denoting an average over time,
\begin{equation}
\overline{s(t_j) Y_N(t_j)} =
\lim_{M\rightarrow\infty}\frac{1}{M}\sum_{j=1}^M s(t_j) Y_N(t_j).
\end{equation} 
Maximizing $C_1$ corresponds to maximizing the coherence between
$s(t_j)$ and $Y_N(t_j)$, namely, it is equivalent to maximizing
information transmission through the devices in Fig. \ref{fig:N-array}.

At first we discuss the numerator of Eq. (\ref{eq:def-C1}).
For large $N$ $Y_N(t_j)$ asymptotically tends to $\langle Y_N(t_j)
\rangle$ according to the law of large number. From
Eq. (\ref{eq:linearization}) the power norm $C_0$ is calculated as,
\begin{equation}
C_0 \approx \frac{G(\lambda)}{\sqrt{D}}||s||^2,
\label{eq:gain}
\end{equation}
where $||s||$ is define as $\sqrt{\overline{s(t_j)^2}}$, namely, the
power norm is proportional to the gain $G(\lambda)$ for a given input
signal.

Next we consider the denominator of Eq. (\ref{eq:def-C1}). For the
purpose we introduce 
$\eta(t_j) \equiv \langle Y_N(t_j) \rangle - Y_N(t_j)$ we have
$\langle \eta(t_j) \rangle = 0$ and 
\begin{eqnarray}
\nonumber
\Delta_j \equiv \langle \eta^2(t_j) \rangle = 
&=& \langle [Y_N(t_j) - \langle Y_N(t_j) \rangle]^2 \rangle \\
\nonumber
&=& \langle Y_N^2(t_j) \rangle - \langle Y_N(t_j) \rangle^2 \\
&=& \frac{1}{N}\Bigl\{P_+(t_j)+P_-(t_j) - \bigl(P_+(t_j)-P_-(t_j)\bigr)^2 \Bigr\}.
\label{eq:Delta}
\end{eqnarray}
$Y_N(t_j)$ is distributed around the ensemble average 
$\langle Y_N(t_j) \rangle$, and $\sqrt{\Delta_j}$ is of order of 
$N^{-1/2}$. Substituting Eqs. (\ref{eq:probability1}),
(\ref{eq:probability2}) and (\ref{eq:probability3}) into
Eq. (\ref{eq:Delta}) yields 
\begin{eqnarray}
\nonumber
\lefteqn{\Delta_j(N,\lambda)} \\
\nonumber
&=& \frac{1}{N}\Bigl\{
\frac{1}{2}\Bigl(
\mbox{erfc}\bigl(\frac{\lambda-\zeta(t_j)}{\sqrt{2}}\bigr)
+\mbox{erfc}\bigl(\frac{\lambda+\zeta(t_j)}{\sqrt{2}}\bigr)
\Bigr) \\
&&-\frac{1}{4}\Bigl(
\mbox{erfc}\bigl(\frac{\lambda-\zeta(t_j)}{\sqrt{2}}\bigr)
-\mbox{erfc}\bigl(\frac{\lambda+\zeta(t_j)}{\sqrt{2}}\bigr)
\Bigr)^2
\Bigr\}.
\end{eqnarray}
Expanding the variance $\Delta_j$ around $\zeta(t_j)=0$ we have
\begin{eqnarray}
\nonumber
\lefteqn{\Delta_j(N,\lambda)}\\
&=& \frac{1}{N}\Bigl\{\mbox{erfc}\Bigl(\frac{\lambda}{\sqrt{2}}\Bigr)-
\frac{1}{\pi}\exp(-\lambda^2)\zeta(t_j)^2\Bigr\} + 
O\Bigl(\zeta(t_j)^4\Bigl).
\label{eq:delta}
\end{eqnarray}
The first term of Eq. (\ref{eq:delta}) results from a fluctuation of the
output signal $Y_N(t_j)$ without the input signal. 

Now we consider $\overline{(Y_N(t_j) - \overline{Y_N(t_j)})^2}$,
calculated as follows~\cite{Chialvo:97}. We have
$\overline{(Y_N(t_j) - \overline{Y_N(t_j)})^2} = \overline{Y_N(t_j)^2}
 \overline{Y_N(t_j)}^2$. 
Since we consider a zero-mean input signal $\overline{Y_N(t_j)} =
0$. From $Y_N(t_j) = \langle Y_N(t_j) \rangle + \eta(t_j)$ we have
\begin{eqnarray}
\nonumber
\overline{Y_N(t_j)^2} 
&=& \overline{\langle Y_N(t_j) \rangle^2} + 2\overline{\langle
  Y_N(t_j) \rangle \eta(t_j)} + \overline{\eta(t_j)^2} \\
&=& \overline{\langle Y_N(t_j) \rangle^2} + \overline{\Delta_j},
\label{eq:overbar-YN2}
\end{eqnarray}
where we use $\overline{\langle Y_N(t_j) \rangle
\eta(t_j)}=0$, which is proven by employing an ergodic assumption. 
Hence from Eqs. (\ref{eq:power-norm}) and
  (\ref{eq:overbar-YN2}) $C_1$ is expressed by 
\begin{eqnarray}
\nonumber
C_1 &=&
\frac{G(\lambda)||s||^2}
{||s||\sqrt{D(\overline{\langle Y_N(t_j) \rangle^2} +
    \overline{\Delta_j})}} \\
\nonumber
&=& 
\frac{1}{\sqrt{1+\frac{\overline{\Delta_j}}{\overline{\langle Y_N(t_j) \rangle^2}}}} \\
&=& \frac{1}{\sqrt{1+c_1^{-1}}},
\label{eq:C1}
\end{eqnarray}
where $c_1$ denotes the ratio between $\overline{\langle Y_N(t_j)
  \rangle^2}$ and the variance $\overline{\Delta_j}$:
\begin{equation}
c_1 = \frac{\overline{\langle
    Y_N(t_j)\rangle^2}}{\overline{\Delta_j}}.
\end{equation}
This statistical measure, which is dimensionless and independent of
scale, is the squared reciprocal of the coefficient of variation.
High $c_1$ indicates low variability of the output signal. 
If $\langle Y_N(t_j) \rangle$ is approximated by the first term of
Eq. (\ref{eq:linearization}), and $\overline{\Delta_j}$ the first term of 
Eq. (\ref{eq:delta}), we obtain
\begin{equation}
c_1 = \frac{2N}{\pi}\frac{\exp(-\lambda^2)}{\mbox{erfc}\Bigl(\frac{\lambda}{\sqrt{2}}\Bigr)}||\zeta||^2,
\label{eq:c1}
\end{equation}
where $||\zeta|| = ||s||/\sqrt{D}$. From Eq. (\ref{eq:C1})
it is clear that $C_1$ is maximized when $c_1$ is maximize. 
Moreover from Eq. (\ref{eq:c1}) it is easily confirmed that $c_1$ is
maximized at $\lambda \approx 0.63$, so that $0.63$ is the optimal
threshold.

In order to infer the input signal $s(t_j)$ from the output signal
$Y_N(t_j)$ it is necessary that $\langle Y_N(t_j) \rangle$ is
sufficiently larger than the fluctuation of $Y_N(t_j)$. If we impose
the condition $c_1>1$, i.e., $C_1 > 1/\sqrt{2}$ we have 
\begin{equation} 
||\zeta|| > \sqrt{\frac{\pi}{2N}}\sqrt{\mbox{erfc}\Bigl(\frac{\lambda}{\sqrt{2}}\Bigr)}\exp\Bigl(\frac{\lambda^2}{2}\Bigr). 
\label{eq:low}
\end{equation}
This inequality assures that we can infer the input signal $s(t_j)$ from
$\frac{\sqrt{D}Y_N(t_j)}{G(\lambda)}$ for $C_1 > 1\sqrt{2}$.

\section{Numerical simulations}
\label{sec:simulation}
\begin{figure}[htb]
\centering
\epsfxsize=250pt
\epsfbox{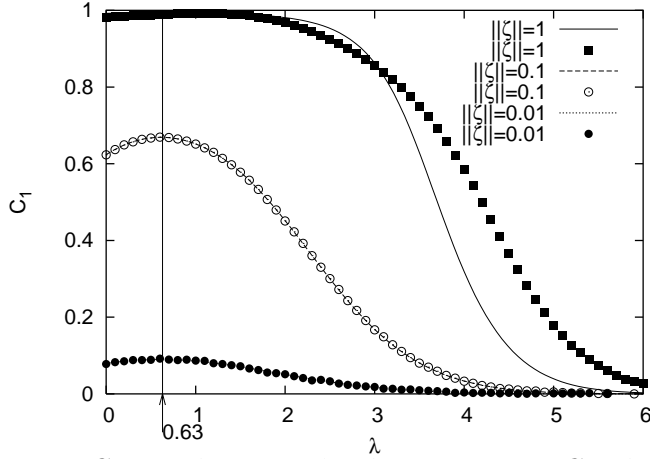}
\caption{The normalized power norm $C_1$, drawn as a
  function of the threshold value $\lambda$ at $N=100$ and a fixed
  amplitude of the input signal. We performed numerical simulations
  using the input signal given by $s(t_j)
  = 0.5 A\sin(2\pi f t_j) + A\cos(4\pi f t_j) + 0.25 A\sin(8\pi f
  t_j)$ at $f = 1.0$ and $T=0.001$. Then we have
  $\frac{||s||}{\sqrt{D}} = ||\zeta|| =
  \sqrt{\frac{21}{32}}A/\sqrt{D}$. We calculate $C_1$ for various $\lambda$ at
  $||\zeta||=1$, $0.1$ and $0.01$. A solid curve represents the
  theoretical relation, Eq. (\ref{eq:C1}) at $||\zeta||=1$, a dashed
  curve at $D=0.1$, a dotted curve at $0.1$ and a dashed curve at $0.01$. 
  Filled squares are results of the numerical simulations at 
  $||\zeta||=1$, unfilled circles at $0.1$, and
  filled circles at $0.01$. It is found that for $||\zeta||=0.1$ and
  $0.01$ the value of $\lambda$ maximizing $C_1$ is $0.63$.  
}
\label{fig:C1}
\end{figure}

Figure \ref{fig:C1} displays the normalized power norm $C_1$ drawn as a
function of $\lambda$ at fixed
$||\zeta||$ from direct numerical simulations of the array of the bithreshold
elements at $N=100$. The input signal is given by $s(t_j) = 0.5
A\sin(2\pi f t_j) + A\cos(4\pi f t_j) + 0.25 A\sin(8\pi f t_j)$ at $f
= 10.0$ and $T=0.001$, where $||\zeta||$ is given by
$\sqrt{\frac{21}{32}}A/\sqrt{D}$. The points are obtained from the
numerical simulations for various $\lambda$ at $||\zeta||=1$, $0.1$
and $0.01$, respectively. The curves represent Eq. (\ref{eq:C1}) at
the same parameters as the numerical simulations. It is found that the
results from the numerical simulations is well fitted by the theoretical
relation for $||\zeta||=0.1$ and $0.01$. $C_1$ has it maximum at 
$\lambda=0.63$. 

However for $||\zeta||=1$ it differs from the theoretical
equation. The reason is because the output signal $Y_N(t_j)$ is not
well approximated by the linear response of the input signal $s(t_j)$
due to the limit of applying the linear response theory. This
disagreement is originated from difference between the nonlinear
response of the system and the linear response assumed in
Sec. \ref{sec:theory}.

\begin{figure}[htb]
\centering
(a)
\epsfxsize=120pt
\epsfbox{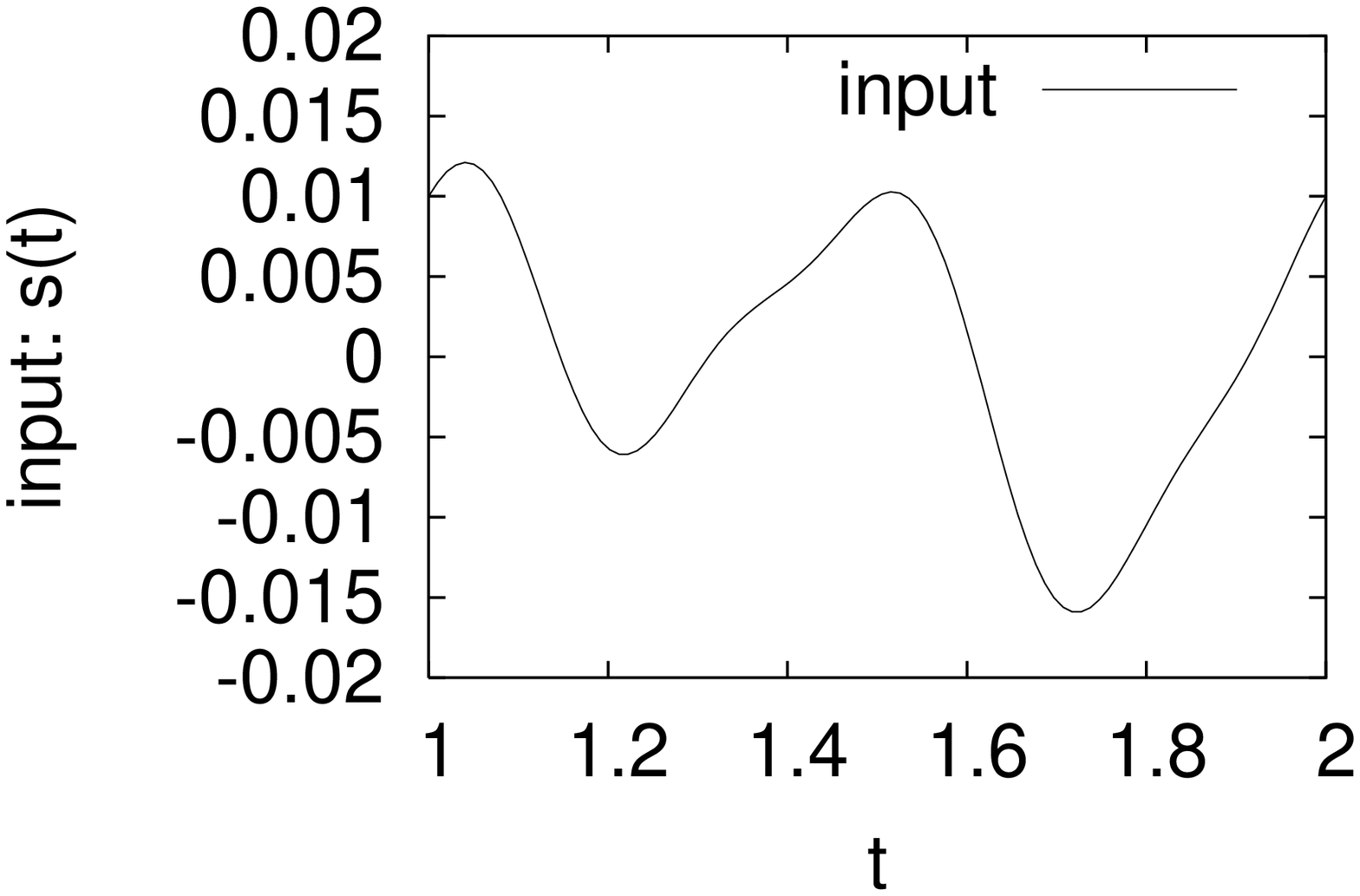}
(b)
\epsfxsize=120pt
\epsfbox{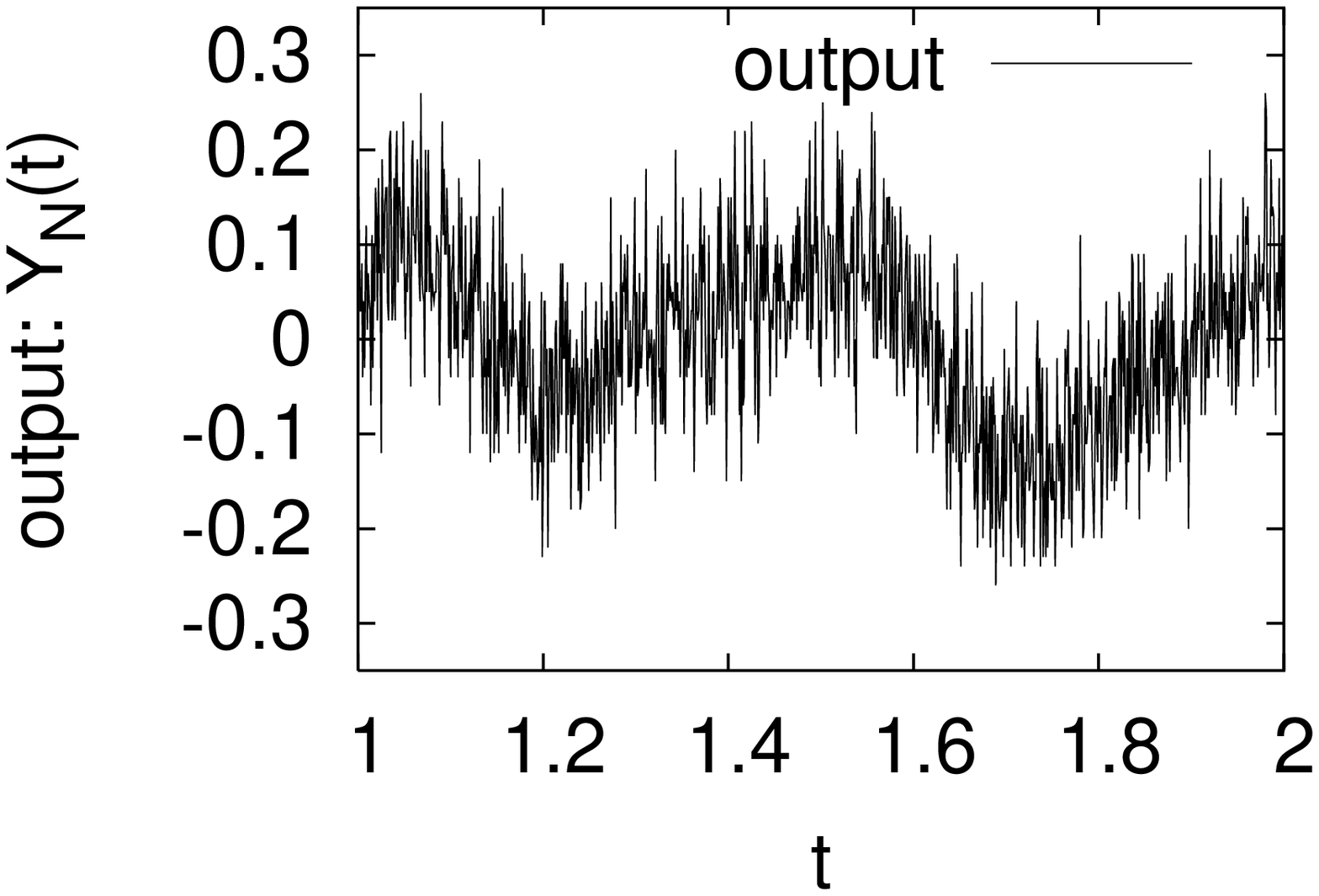}\\
(c)
\epsfxsize=120pt
\epsfbox{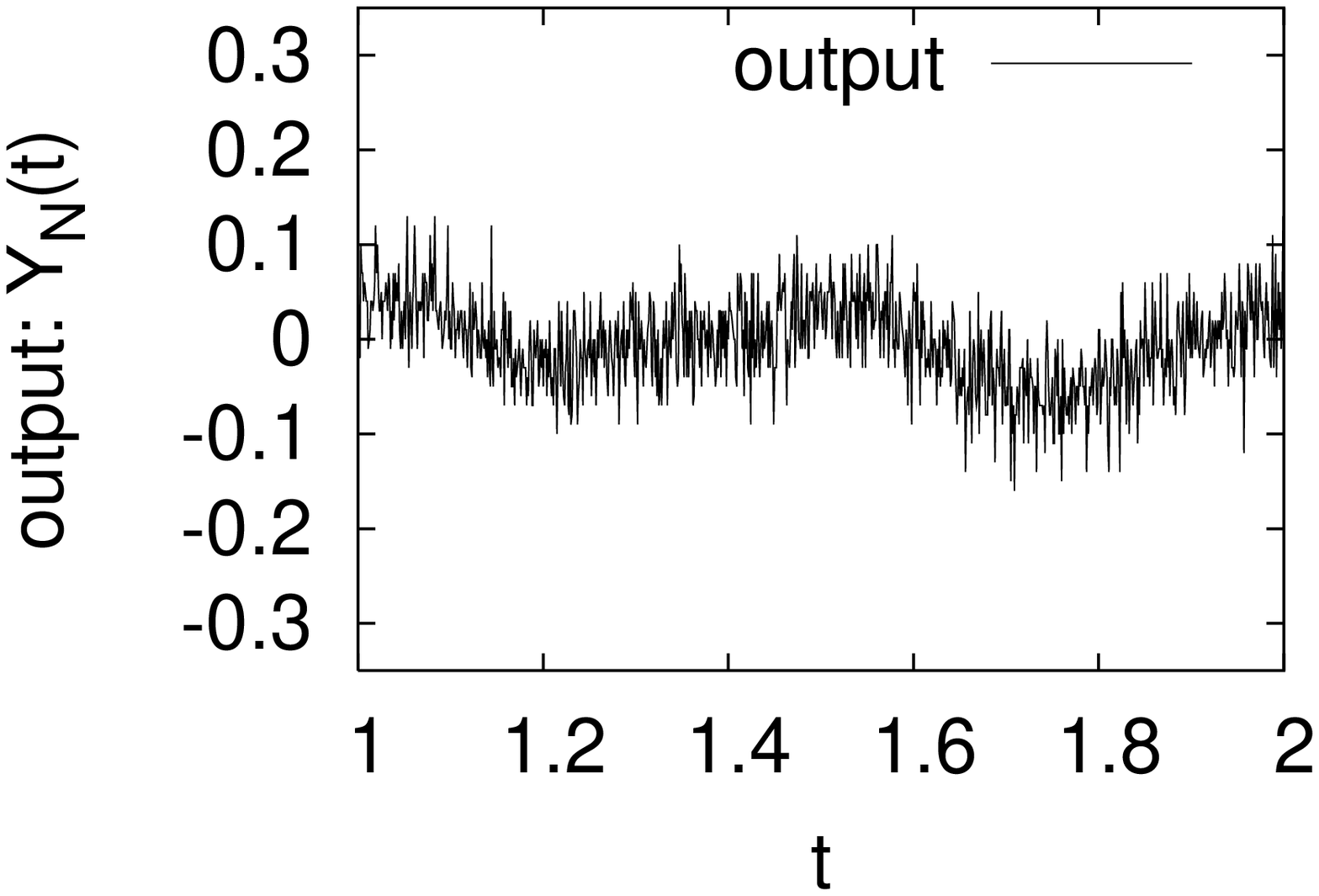}
(d)
\epsfxsize=120pt
\epsfbox{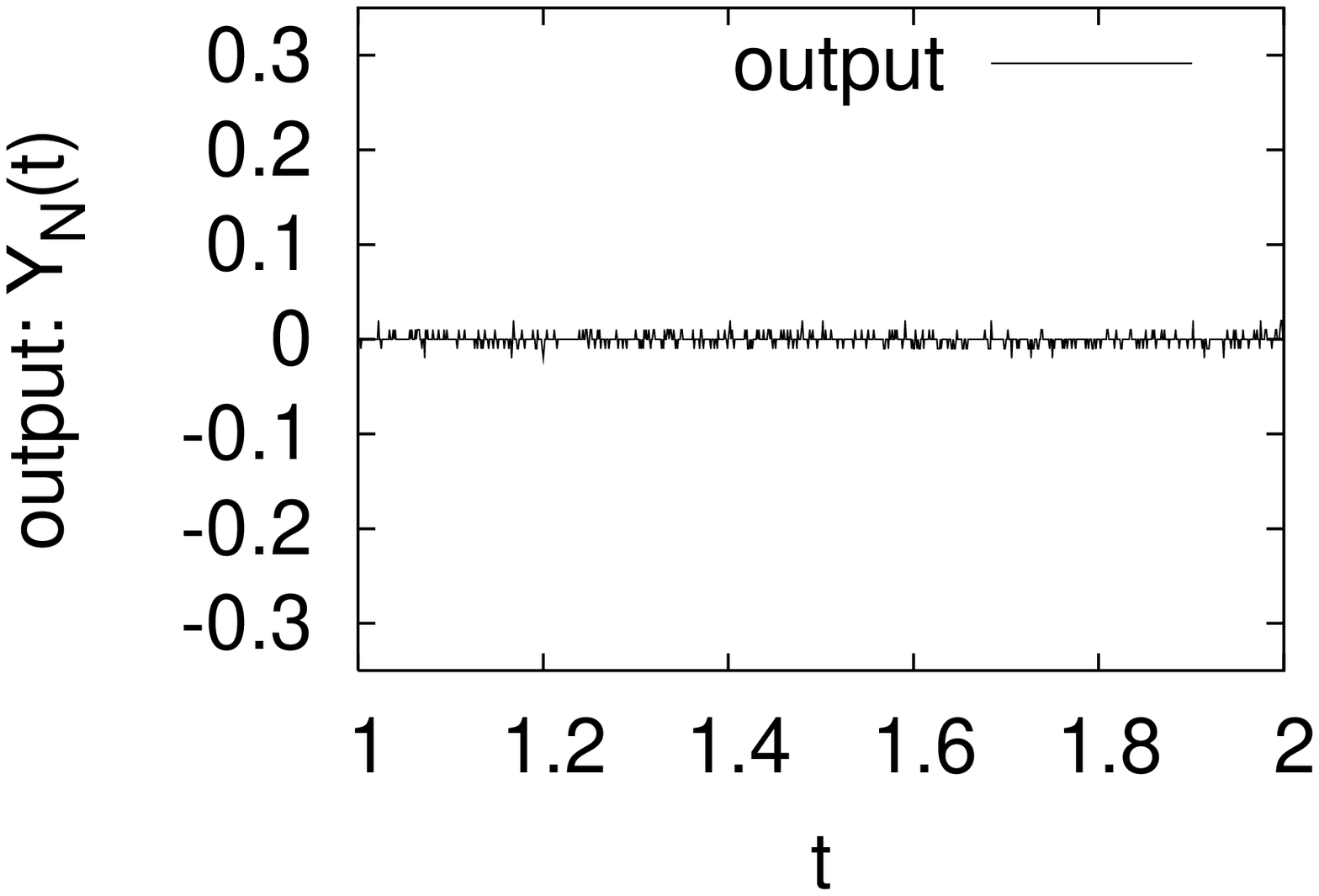}
\caption{Time series of the input signal $s(t_j)$, of which the
  waveform is the same as Fig. \ref{fig:C1} (a). We performed
  the numerical simulation at $N=100$, $D=0.1$ and $||\zeta||=0.1$. 
  The output signal $Y_N(t_j)$ at $\lambda = 0.63$ ($C_1=0.669$) (b),
  at $\lambda = 1.5$ ($C_1=0.577$) (c) and $\lambda = 3.0$
  ($C_1=0.166$) (d).}
\label{fig:output}
\end{figure}

We demonstrate the output signal $Y_N(t_j)$ for various $\lambda =
0.63$, $1.5$ and $3.0$ as shown in Fig. \ref{fig:output}. The output
signal $Y_N(t_j)$ is similar to the input signal $s(t_j)$ in the
order for $\lambda$ shown. In this demonstration the input signal is
given by a periodic signal. Naturally the theoretical equation of $C_1$
which we obtained is applicable to any input signal (of course
an aperiodic signal) satisfied with $||\zeta|| < 1$. Specifically when
the input signal is satisfied with Eq. (\ref{eq:low}) the output
signal gives a good approximation of the input signal at $\lambda = 0.63$. 

\section{Conclusion}
\label{sec:conclusion}
We have investigated the parallel array of bithreshold elements 
both theoretically and numerically. We give an analytical
approximation of the normalized power norm $C_1$ under the assumption
that the norm of the input signal $s(t_j)$ is smaller than the
standard deviation of the additive noises without distinguishing
between subthreshold input and suprathreshold input. We confirmed that the
theoretical approximation of $C_1$ is consistent with the results
obtained from the direct numerical simulations of the array of the
bithreshold elements when the norm of the input signal is smaller than
the standard deviation of the additive noises. While for the larger
norm than the standard deviation the difference between the
approximation and the numerical results appears. This disagreement is
originated from the nonlinear response of the system. We demonstrated
that the output signal gives a good approximation of the input signal
at an appropriate threshold. We clarify that the optimal threshold,
where the normalized power norm has a maximum value, is given by 0.63
times the standard deviation of the noises. 

Our study may be applied to a sophisticated array of amplifiers. Moreover
the result shows that a collection of simple bithreshold sensors can
detect a weak signal under an independently noisy environment.

%=============================================================================
%=============================================================================
\newpage
%==============================================================================
%==============================================================================
\newpage
%==============================================================================
%=============================================================================

%\end{multicols}
\end{document}